\documentclass[12pt]{article}
\usepackage{amsfonts,amsmath}
\usepackage[english]{babel}

\newcommand{\al}{\alpha}
\newcommand{\gb}{\beta}
\newcommand{\gga}{\gamma}
\newcommand{\gl}{\lambda}
\newcommand{\gk}{\kappa}
\newcommand{\dal}{\dot \alpha}
\newcommand{\dgb}{\dot \beta}
\newcommand{\dgga}{\dot \gamma}
\newcommand{\q}{\,,\qquad}
\newcommand{\p}{\partial}
\newcommand{\f}{\frac}
\newcommand{\fO}{\mathbf{\Omega}}
\newcommand{\fh}{\mathbf{h}}
\newcommand{\D}{\mathcal{D}}
\newcommand{\G}{\mathcal{G}}
\newcommand{\fD}{\mathcal{D}}
\newcommand{\fV}{\mathcal{V}}
\newcommand{\M}{\mathcal{M}}
\newcommand{\be}{\begin{equation}}
\newcommand{\ee}{\end{equation}}

\makeatletter \@addtoreset{equation}{section} \makeatother

\renewenvironment{thebibliography}[1]
  { {\bf \LARGE References} \begin{list}{[\arabic{enumi}]}
    {\usecounter{enumi} \setlength{\parsep}{0pt}
     \setlength{\itemsep}{0pt} \settowidth{\labelwidth}{#1.}
     \sloppy
    }}{\end{list}}

\begin{document}

\begin{flushright}
FIAN/TD/13-08\\
\end{flushright}

\vspace{0.5cm}

\begin{center}
{\large\bf Unfolded Description of $AdS_4$ Kerr Black Hole}

\vspace{1 cm}

{\bf V.E.~Didenko, A.S.~Matveev and M.A.~Vasiliev}\\
\vspace{0.5 cm} {\it
I.E. Tamm Department of Theoretical Physics, Lebedev Physical Institute,\\
Leninsky prospect 53, 119991, Moscow, Russia }

\vspace{0.6 cm}
didenko@lpi.ru, matveev@lpi.ru, vasiliev@lpi.ru \\
\end{center}

\vspace{0.4 cm}

\begin{abstract}
\noindent It is shown that $AdS_4$ Kerr black hole is a solution of
simple unfolded differential equations that form a deformation of
the zero-curvature description of empty $AdS_4$ space-time. Our
construction uses the Killing symmetries of the Kerr solution. All
known and some new algebraic properties of the Kerr-Schild solution result
from the obtained black hole unfolded system in the coordinate-independent
way. Kerr-Schild type solutions of free equations in $AdS_4$ for
massless fields of any spin associated with the proposed black hole
unfolded system are found.

\end{abstract}

\section{Introduction}

Since its discovery in 1963, the celebrated Kerr metric \cite{Kerr}
is still a hot research topic\footnote{For the history of question
we refer to the recent publication by Roy Patric Kerr
\cite{Kerr-rev}.}. Being of great physical significance, it exhibits
deep mathematical beauty. In particular, the Kerr-Schild form
\cite{K-S} of the Kerr metric illustrates the remarkable fact that
Kerr black hole (BH), being an exact solution of Einstein equations,
also verifies its first order fluctuational part, {\it i.e.,} the
second order terms vanish independently.  In other words, Kerr BH
solves free spin-two equations. This property suggests the relevance
of perturbative study of corrections to Kerr-like solutions in such
field-theoretical extensions of gravity as string theory or
higher-spin (HS) gauge theory (see \cite{QG,Gol,sor,SSS,solv} for
reviews). The Kerr-Schild form allowed Myers and
Perry to generalize the four-dimensional Kerr metric to any
dimension \cite{M-P}.

BHs with non-zero cosmological constant have also been
extensively studied in the context of $AdS/CFT$ correspondence
\cite{malda, witt,hawk-hun,chamb} where the (anti)-de Sitter
geometry plays a distinguished role. Although the generalization
of the  Kerr solution that describes a rotating BH in
$AdS_4$ was discovered by Carter
\cite{Car} soon after Kerr's paper \cite{Kerr}, its higher
dimensional generalization has been found only recently. A
generalization of the Myers-Perry solution with non-zero
cosmological term in five dimensions has been given by Hawking,
Hunter and Taylor-Robinson in \cite{Hawk} and then extended to any
dimension by Gibbons, Lu, Page and Pope in \cite{Gib}. These
higher dimensional BHs are shown \cite{Frolov} to possess
the hidden symmetry associated with the Yano-Killing tensor
\cite{Yano} which is the characteristic feature of Petrov type
$D$ metrics.

Despite considerable progress in the construction of different
BH metrics, including the charged rotating Kerr-Newman
solution \cite{Newman}, the generalization of $AdS_4$
Taub-NUT to any dimension \cite{Gibon2} and discovery of various
supergravity BHs in the series of works by Chong, Cvetic, Gibbons, Lu,
Pope \cite{Cv1,Cv3}, a number of problems
remain open. In particular, a generalization of the well-known four-dimensional
 Kerr-Newman solution to charged rotating BH
solution in any dimension is available neither in flat nor in
$(A)dS$ case. This
suggests that some more general approaches can be useful.

Usually an analysis in the BH background uses
a particular coordinate system. The aim of this paper is to
reformulate the $AdS_4$ Kerr BH in a
coordinate-independent way using the formalism of the unfolded
dynamics \cite{Ann, act} that operates with first-order covariant
field equations. We show how the $AdS_4$ Kerr BH arises
as a solution of the BH unfolded system (BHUS)
associated with a certain BH Killing vector.

In general, unfolded equations generalize the Cartan-Maurer equations for
 Lie algebras by incorporating $p$-form gauge potentials. They
provide a powerful tool for the study of partial differential equations.
More precisely, consider, following \cite{Ann}, a set of
differential $p$-forms $W^A(x)$ with $p \ge 0$ and generalized
curvatures $R^A$ defined as
\be\label{def1}
R^A = dW^A + F^A(W)\,,
\ee
where $d = dx^n\frac{\partial}{\partial x^n}$ is the
space-time exterior differential and some functions $F^A$ are built of
wedge products of
the differential forms $W^B$. The functions $F^A(W )$ subjected the
generalized Jacobi identity
\be\label{def2}
F^B \frac{\delta F^A}{\delta W^B} = 0
\ee
guaranty the generalized Bianchi identity
$
dR^A =R^B \frac{\delta F^A}{\delta W^B}\,
$
which shows, in particular, that the differential equations for
$W^A(x)$
\be\label{unfo}
R^A = 0
\ee
are consistent. In this case the equations
(\ref{unfo}) are called unfolded. Note that every unfolded system
is associated with some solution of the generalized Jacobi identity
(\ref{def2}) on the functions $F^A(W)$ built of exterior products
of the differential forms $W^A$ which, in turn, defines a
free differential algebra \cite{FDA1,FDA2} (more
precisely, its generalization to the case with zero-forms included).

The reformulation of partial differential equations in the
unfolded form has a number of advantages. In particular,
the first-order equations formulated in terms of exterior
algebra are manifestly coordinate-independent and reconstruct
a local solution in terms of its values at any given point of
space-time modulo gauge invariance for $p$-forms with $p\geq 1$.
(For the recent summary of the properties of unfolded dynamics we refer
the reader to \cite{33}.)

As an example of the unfolded system
consider the Killing equation in the $AdS_4$ space written in the
local frame
\be\label{exmp}
DK_{a}=\kappa_{ba}h^{b}\,,\quad D\kappa_{ab}=2\gl^2K_{[a}h_{b]}\,,\quad
D^2=R_{ab}=\gl^2h_{a}\wedge h_{b}\,,
\ee
where $K_a$ is some Killing vector, $\kappa_{ab}=-\kappa_{ba}$, $D$ is the
Lorentz covariant differential, $h_{a}$ is the $AdS_4$ vierbein
one-form, $-\gl^2$ is the cosmological constant, and $a,b=0, \dots, 3$. The $AdS_4$
geometry is described by the zero curvature equations
\be\label{zero-1}
d\Omega_{ab}+\Omega_{a}{}^{c}\wedge\Omega_{cb}-\gl^2 h_{a}\wedge h_{b}=0\,,
\ee
\be\label{zero-2}
dh_{a}+\Omega_{a}{}^{c}\wedge h_{c}=0\,,
\ee
where $\Omega_{ab}$ is the $AdS_4$ Lorentz connection one-form.
The first of the equations
\eqref{exmp} is obviously equivalent to the $AdS_4$ Killing
equation as it implies
$
D_{a}K_{b}+D_{b}K_{a}=0,
$
while the second one follows as a consequence of Bianchi identities
for the space of constant curvature. The equations
\eqref{exmp}--\eqref{zero-2} represent the unfolded system for the
set of forms 
$W_{A}=(h_{a}, \Omega_{ab}, K_{a}, \kappa_{ab}).$

In this paper we show that there exists a simple deformation of the
equations \eqref{exmp}--\eqref{zero-2} with the same set of forms
that describes the $AdS_4$ Kerr BH in four dimensions with the BH
mass being a free deformation parameter.
Since the proposed BH Unfolded System (BHUS)
expresses all derivatives of fields in terms of the fields
themselves, all invariant relationships on the derivatives
of BH geometric quantities become direct consequences
of BHUS. In particular, it  manifestly expresses the Petrov type
$D$ BH
Weyl tensor in terms of the Killing two-form.

The $AdS_4$ Killing two-form that results from the exterior differential
 of the one-form dual to the  Killing vector plays the central role in our
consideration.  The integrability properties of the $AdS_4$ BHUS  imply,
in particular, the existence of the sourceless Maxwell tensor and Yano-Killing
tensor, both being proportional to the Killing
two-form. Let us stress that, even though the Killing two-form
in $AdS_4$ has non-vanishing
divergence, it remains related to the BH Weyl tensor and to the
sourceless Maxwell field.

Although the obtained results may have different applications,
the original motivation for this study was due to the search
 of the Kerr BH  solution in
the nonlinear HS gauge theory which itself is formulated in the
unfolded form and $AdS$ space
(see, e.g., reviews \cite{Gol,solv} and references
therein).  The proposed formulation looks particularly promising
in that respect. We show how
 the proposed BHUS makes it possible to
generate Kerr-Schild like solutions of the free bosonic HS
field equations for all spins $s=0, 1\dots$ in $AdS_4$. Remarkably,
for $s=1$ and $s=2$ the obtained solutions satisfy in addition  the
minimally coupled to gravity equations. For other spins
the covariantization of Fronsdal equations \cite{CF} in the Kerr-Schild
metric leads to certain nonlinear terms which are explicitly
calculated.

The rest of the paper is organized as follows. First, in
Section \ref{genfacts} we recall the basic facts about the $AdS_4$ Kerr BH
solution, its Kerr-Schild metric decomposition and its Killing symmetry.
In Section \ref{cart} we reconsider the Cartan formalism most appropriate for
our purpose.  In Section \ref{FDA} we present the BH unfolded system,
build the null geodesic congruence using Killing projectors, find relations
between the Killing two-form, Maxwell tensor and Yano-Killing tensor.
In Section \ref{Kerr} we reproduce the Kerr solution from $AdS_4$ space
via a Kerr-Schild type algebraic shift. Finally, in Section \ref{HS} we
show how Kerr-Schild type solutions for free HS equations can be obtained.
Section \ref{sum} contains summary and conclusions.

\section{$AdS_4$ Kerr black hole} \label{genfacts}

A rotating BH of mass $M$ in $AdS_4$ admits the
Kerr-Schild form\footnote{Throughout this paper we use units with
$8\pi G=1$.} \cite{Car}
\be\label{ks}
g_{mn}(X)=\eta_{mn}(X)+\f{2M}{U(X)}k_m(X)
k_{n}(X)\,,\quad
g^{mn}(X)=\eta^{mn}(X)-\f{2M}{U(X)}k^m(X)
k^n(X)\,,
\ee
where $\eta_{mn}(X)$ ($m,n=0,\ldots, 3$) is the background $AdS_4$ metric with
negative cosmological constant $-\gl^2$  and $k^m(X)$ defines the
null geodesic congruence with respect to both full metric
$g_{mn}(X)$ and the background one $\eta_{mn}(X)$
\be \label{congr-def}
k^{m}k_m=0\,,\quad k^{m}\fD_{m}k_n=k^m D_m
k_n=0\,.
\ee
Here $\fD$ and $D$ are full and background covariant differentials, respectively.

A useful coordinate system has the background metric of the form \cite{Gib}
\be\label{fon}
\eta^{mn}= \left(
\begin{array}{cccc}
  \frac{1-a^2\lambda^2}{(1+r^2\lambda^2)(1-\lambda^2 \frac{a^2z^2}{r^2})} & 0 & 0 & 0 \\
  0 & -1-\lambda^2(x^2-a^2) & -\lambda^2 xy &  -\lambda^2 xz\\
  0 & -\lambda^2 xy & -1-\lambda^2(y^2-a^2) & -\lambda^2 yz\\
  0 & -\lambda^2 xz & -\lambda^2 yz & -1-\lambda^2 z^2\\
\end{array}
\right)\,,
\ee
where the radial coordinate $r(X)$ is defined through the ellipsoid
of revolution equation
\be\label{ellips}
\f{x^2+y^2}{r^2+a^2}+\f{z^2}{r^2}=1
\ee
and $a$ is a rotational parameter. The BH has angular
momentum $J=Ma$.

Components of the Kerr-Schild vector $k^m(X)$ are
\be\label{k-comp}
k^0=\frac{1}{1+r^2\lambda^2}, \quad k^1=-\frac{xr-ay}{r^2+a^2},
\quad k^2=-\frac{yr+ax}{r^2+a^2}, \quad k^3=-\frac{z}{r}\,,
\ee
and
\be \label{U-def}
U(X)=r+\f{a^2z^2}{r^3}\,.
\ee
Direct calculation gives
\be\label{dk}
\f{2}{U}=\frac{1}{Q}+\frac{1}{\bar Q}=-D_{m}k^{m}=-\fD_{m}k^{m}\,,
\ee
where
\be \label{Qdef}
Q=r- \frac{iaz}{r}, \qquad \bar Q=r+ \frac{iaz}{r}\,.
\ee

Note that the metric \eqref{ks} still provides a Kerr solution after
the transformation
\be\label{tau}
\tau(a,x,y,z,t)=(-a,-x,-y,-z,t)\,.
\ee
In other words, the Kerr-Schild Ansatz
also works for the vector $n^i = \tau(k^i)$.

One can check that the Maxwell tensor
\be\label{Max-tensor}
F=dA^{(1)}=dA^{(2)}\,,\quad A^{(1)}_{m}=\f{k_{m}}{U},
\quad A^{(2)}_{m}=\f{n_{m}}{U}
\ee
 verifies the sourceless Maxwell equations  both in the $AdS_4$
and in the BH geometry
\be
D_{m}F^{m}{}_n=\fD_{m}F^{m}{}_n=0\,.
\ee

The Kerr BH has two Killing vectors, time translation $\fV_t^m$ and rotation around $z$-axis
$\fV_\phi^m$
\be\label{t-kill}
\fV_t^m = \f{\p}{\p t}=(1,0,0,0), \quad
\fV_\varphi^m = \f{\p}{\p\varphi}=(0, y, -x, 0).
\ee

Let us introduce a Killing two-form $\kappa = d K$ as the exterior derivative
of the one-form $dx^m K_m$ dual to some Killing vector $K^m$.
This two-form will also be called Papapetrou field. It was originally
introduced in \cite{Papa}, where it was shown to give rise to the
sourceless Maxwell tensor in  Ricci flat manifolds with
isometries.

A particular linear combination $\fV^{m}$ of the Killing vectors
\eqref{t-kill}
\be\label{Kill}
\fV^m=(1, a\gl^2 y, -a\gl^2 x, 0)
\ee
that satisfies the condition $k_m \fV^{m}=1$, will be used later
on for the definition of the Killing two-form associated with
 a Kerr BH in $AdS_4$.

\section{Cartan formalism} \label{cart}

Let $dx^m \fO_{m}{}^{ab}$ be an antisymmetric one-form Lorentz
connection and $dx^m \fh_m{}^a$ be a vierbein one-form.
These can be identified with the  gauge fields of the $AdS_4$
symmetry algebra $o(3,2)$. The corresponding $AdS_4$ curvatures
have the form
\be \label{1Cartan}
\mathbf{R}^{ab} = d\fO^{ab} + \fO^{ac}\wedge\fO_{c}{}^{b}
-\lambda^2 \fh^a \wedge \fh^b\,,
\ee
\be \label{2Cartan}
\mathbf{R}^a = d\fh^a + \fO^{ac}\wedge \fh_c\,,
\ee
where  $a,b,c=0,\dots, 3$ are Lorentz indices.

The zero-torsion condition $\mathbf{R}^a =0$ expresses
algebraically the Lorentz connection $\fO$ via derivatives of
$\fh$. Then the $\lambda$-independent part of the curvature two-form
\eqref{1Cartan} identifies with the Riemann tensor. Einstein
equations imply that the Ricci tensor vanishes up to a constant
trace part proportional to the cosmological constant. In other
words, only those components of the tensor \eqref{1Cartan} may
remain non-vanishing on-shell that belong to the Weyl tensor
\be \label{Cartan-Weyl}
\mathbf{R}_{ab} = \frac12 \fh^c \wedge \fh^d C_{cdab}\,,
\ee
where $C_{abcd}$ is the Weyl tensor in the local frame,
$
C_{abcd}=-C_{bacd}=-C_{abdc}=C_{cdab}.
$

The analysis in four dimensions considerably simplifies in
spinor notation.
Vector notation are translated  to the spinor one and vice versa
with the help of Pauli $\sigma$-matrices. For example,
for a Lorentz vector $U_a$ we have
\be
U_{\alpha\dot{\alpha}}=(\sigma^a)_{\alpha\dot{\alpha}}U_a\,,\quad
U_a =
\frac12(\sigma_a)^{\alpha\dot{\alpha}}U_{\alpha\dot{\alpha}}\,.
\ee
Spinor indices are raised and lowered  by the $sp(2)$ antisymmetric
tensors $\varepsilon_{\alpha\beta}$ and
$\varepsilon_{\dot\alpha\dot\beta}$
\be
\xi_{\alpha}=\xi^{\beta}\varepsilon_{\beta\alpha}\,,\quad
\xi^{\alpha}=\varepsilon^{\alpha\beta}\xi_{\beta}\q
\bar\xi_{\dal}=\bar\xi^{\dgb}\varepsilon_{\dgb\dal}\,,\quad
\bar\xi^{\dal}=\varepsilon^{\dal\dgb}\bar\xi_{\dgb}\,,
\ee
where $\varepsilon_{12}=\varepsilon^{12}=1$,
$\varepsilon_{\alpha\beta}=-\varepsilon_{\beta\alpha}$, $\varepsilon^{\alpha\beta}=-\varepsilon^{\beta\alpha}$.

Lorentz irreducible spinor decompositions of the Maxwell and Weyl tensors
$F_{ab}$ and $C_{abcd}$ read, respectively, as
\be \label{sp2decomp}
F_{\alpha\dot{\alpha}\beta\dot{\beta}} =\varepsilon_{\alpha\beta}
\bar{F}_{\dot{\alpha}\dot{\beta}}+\varepsilon_{\dot{\alpha}\dot{\beta}}
F_{\alpha\beta}\q
C_{\alpha\dot{\alpha}\beta\dot{\beta}\gamma\dot{\gamma}\delta\dot{\delta}}
= \varepsilon_{\alpha\beta}\varepsilon_{\gamma\delta}
\bar{C}_{\dot{\alpha}\dot{\beta}\dot{\gamma}\dot{\delta}}
+\varepsilon_{\dot{\alpha}\dot{\beta}}\varepsilon_{\dot{\gamma}\dot{\delta}}
C_{\alpha\beta\gamma\delta}\,,
\ee
where\footnote{The symmetrization
over repeated spinor indices is implied.}
 $F_{\alpha\alpha}$ and $C_{\alpha\alpha\alpha\alpha}$ are
totally symmetric multispinors.

To describe Killing symmetries let us rewrite the Cartan equations
\eqref{1Cartan}, \eqref{2Cartan} in  spinor notation. Lorentz
connection one-forms $\fO_{\alpha\alpha},
\bar\fO_{\dot\alpha\dot\alpha}$ and vierbein one-form
$\fh_{\alpha\dot\alpha}$ can be identified with the gauge fields
of $sp(4) \sim o(3,2)$.  Vacuum Einstein equations with
cosmological constant acquire the form
\be \label{RHC}
\mathcal{R}_{\alpha\alpha} =  d\fO_{\alpha\alpha} +
\fO_{\alpha}{}^\gamma \wedge \fO_{\gamma\alpha} =
\frac{\lambda^2}{2}\, \mathbf{H}_{\alpha\alpha} +
\frac18 \mathbf{H}^{\gamma\gamma}C_{\gamma\gamma\alpha\alpha}\,,
\ee
\be \label{RHCc}
\mathcal{\bar R}_{\dot\alpha\dot\alpha} =  d\bar{\fO}_{\dot\alpha\dot\alpha} +
\bar{\fO}_{\dot\alpha}{}^{\dot\gamma} \wedge \bar{\fO}_{\dot\gamma\dot\alpha}=
\frac{\lambda^2}{2}\, \mathbf{\bar{H}}_{\dot\alpha\dot\alpha}  +
\frac18 \mathbf{\bar H}^{\dot\gamma\dot\gamma}\bar{C}_{\dot\gamma\dot\gamma\dot\alpha\dot\alpha}\,,
\ee
\be \label{Tfree}
\mathcal{R}_{\alpha\dot{\alpha}} = d\fh_{\alpha\dot{\alpha}} +
\frac12 \fO_{\alpha}{}^{\gamma}\wedge \fh_{\gamma\dot{\alpha}} +
\frac12 \bar{\fO}_{\dot{\alpha}}{}^{\dot{\gamma}}\wedge
\fh_{\alpha\dot{\gamma}}=0\,,
\ee
where $\mathcal{R}_{\alpha\beta},
\bar{\mathcal{R}}_{\dot\alpha\dot\beta}$ are the components of the Loretnz curvature two-form
\be \label{D2Rcal}
\fD^2\xi_{\alpha\dot\alpha} = \frac12 \mathcal{R}_{\alpha}{}^{\beta}\xi_{\beta\dot\alpha}
+ \frac12 \mathcal{\bar R}_{\dot\alpha}{}^{\dot\beta}\xi_{\alpha\dot\beta}
\ee
and
\be \label{Hcap}
\mathbf{H}^{\alpha\alpha} = \fh^{\alpha}{} _{\dot\alpha} \wedge
\fh^{\alpha\dot\alpha}\,, \qquad
\bar{\mathbf{H}}^{\dot\alpha\dot\alpha} =
\fh_{\alpha}{}^{\dot\alpha} \wedge \fh^{\alpha\dot\alpha}\,.
\ee

The Killing equation is
\be \label{1eq}
\fD \fV_{\alpha\dot\alpha} = \frac12
\fh^{\gamma}{}_{\dot\alpha}\kappa_{\gamma\alpha}+ \frac12
\fh_{\alpha}{}^{\dot\gamma}\bar\kappa_{\dot\gamma\dot\alpha}\,,
\ee
where $\kappa_{\alpha\alpha}$ and
$\bar\kappa_{\dot\alpha\dot\alpha}$  just represent non-zero components of first
derivatives of the Killing vector $\fV_{\al\dal}$.
In vector notation this gives
$\D_{m}\fV_{n}=\gk_{mn}=-\gk_{nm}$, hence leading to
$\D_{(m}\fV_{n)}=0$.

Differentiation of
\eqref{1eq} with the help of \eqref{RHC}, \eqref{RHCc} and \eqref{D2Rcal} yields
\begin{eqnarray}
\fD\kappa_{\alpha\alpha} &=&  \lambda^2
\fh_{\alpha}{}^{\dot\gamma}\fV_{\alpha\dot\gamma} +
\frac14 \fh^{\beta\dot\beta}\fV^{\beta}{}_{\dot\beta}C_{\beta\beta\alpha\alpha}\,, \label{2eq}\\
\fD\bar\kappa_{\dot\alpha\dot\alpha} &=& \lambda^2
\fh^{\gamma}{}_{\dot\alpha}\fV_{\gamma\dot\alpha} + \frac14
\fh^{\beta\dot\beta}\fV_{\beta}{}^{\dot\beta}\bar
C_{\dot\beta\dot\beta\dot\alpha\dot\alpha}\,. \label{3eq}\
\end{eqnarray}
with compatibility conditions
$
\fD^{\beta\dot\alpha}C_{\beta\alpha\alpha\alpha} = 0\,, 
\fD^{\alpha\dot\beta}\bar C_{\dot\beta\dot\alpha\dot\alpha\dot\alpha} = 0\,.
$
In case $C_{\al\al\al\al}=0$, $\bar C_{\dal\dal\dal\dal}=0$ the equations
\eqref{1eq}--\eqref{3eq} describe an isometry of
$AdS_4$.

\section{Black hole unfolded equations}\label{FDA}

\subsection{Consistency condition for Killing unfolded system}

Let us investigate solutions of Einstein equations with negative
cosmological constant  and the Weyl tensor of the form
\be \label{weyl}
C_{\alpha\alpha\alpha\alpha} = f(X) \kappa_{\alpha\alpha}\kappa_{\alpha\alpha}\,,
\ee
where $f(X)$ is some function of space-time coordinates $X$, and
$\kappa_{\al\al}$ is a Killing two-form corresponding to some
Killing vector $\fV_{\al\dal}$. So, we assume  at
least one isometry. The Weyl tensor \eqref{weyl}
is of the type $D$ by Petrov classification \cite{Petrov}.

From the Killing equations \eqref{1eq}--\eqref{3eq}
 we derive the following unfolded equations
\begin{eqnarray}
\fD \fV_{\alpha\dot\alpha} & =& \frac12
\fh^{\gamma}{}_{\dot\alpha}\kappa_{\gamma\alpha} +
\frac12 \fh_{\alpha}{}^{\dot\gamma}\bar\kappa_{\dot\gamma\dot\alpha}\,, \label{FDA1} \\
\fD\kappa_{\alpha\alpha} &=&\lambda^2
\fh_{\alpha}{}^{\dot\gamma}\fV_{\alpha\dot\gamma} +
\frac{f}{4} \fh^{\beta\dot\beta}\fV^{\beta}{}_{\dot\beta}\kappa_{\beta\beta\alpha\alpha}\,, \label{FDA2}\\
\fD\bar\kappa_{\dal\dal} &=&\lambda^2
\fh^{\gga}{}_{\dal}\fV_{\gga\dal} +
\frac{\bar f}{4} \fh^{\beta\dot\beta}\fV_{\beta}{}^{\dot\beta}\bar\kappa_{\dgb\dgb\dal\dal}\,, \label{FDA2c}\\
\mathcal{R}_{\alpha\alpha} &=&
\f{\lambda^2}{2}\mathbf{H}_{\alpha\alpha}
+\f{f}{8}\mathbf{H}^{\beta\beta}\kappa_{\beta\beta\alpha\alpha}\label{FDA3}\,,\\
\mathcal{\bar R}_{\dot\alpha\dot\alpha} &=&
\f{\lambda^2}{2}\mathbf{\bar H}_{\dot\alpha\dot\alpha}
+\f{\bar f}{8}\mathbf{\bar H}^{\dot\beta\dot\beta}\bar\kappa_{\dot\beta\dot\beta\dot\alpha\dot\alpha}\label{FDA3c}\,,\\
\fD\fh_{\alpha\dot\alpha}&=&0\,,\\
df &=& -( \frac{1}{12} f^2 + \frac{5f\lambda^2}{2\kappa^2} )
\fh^{\alpha\dot\gamma}\fV^{\alpha}{}_{\dot\gamma}\kappa_{\alpha\alpha}\,,\label{feq}\\
d\bar{f} &=& -( \frac{1}{12} \bar{f}^2 + \frac{5\bar
f\lambda^2}{2\bar\kappa^2} )
\fh^{\gga\dal}\fV_{\gga}{}^{\dal}\bar\kappa_{\dal\dal}\,,\label{feqc}\label{FDA4}
\end{eqnarray}
where
$\mathbf{H}^{\alpha\alpha}$ and $\mathbf{\bar H}^{\dot\alpha\dot\alpha}$ are defined in \eqref{Hcap}, $\mathcal{R}_{\alpha\alpha}$ and $\mathcal{\bar R}_{\dot\alpha\dot\alpha}$
are the curvatures \eqref{RHC} and \eqref{RHCc},
and we use notations
$
\kappa_{\alpha\beta}\kappa^{\beta}{}_{\gamma}=\kappa^2
\varepsilon_{\alpha\gamma}\,,
\gk_{\al\al\al\al}=\gk_{\al\al}\gk_{\al\al}.
$

The system \eqref{FDA1}--\eqref{FDA4} has the unfolded form
\eqref{unfo} and is formally consistent. We call it BH unfolded
system (BHUS). It is invariant under the
following  transformation
\be \label{bhussym} \tau_\mu:
\quad (\fV_{\alpha\dot\alpha}, \kappa_{\alpha\alpha},
\bar\kappa_{\dot\alpha\dot\alpha}, f, \bar f) \rightarrow
(\mu\fV_{\alpha\dot\alpha}, \mu\kappa_{\alpha\alpha},
\mu\bar\kappa_{\dot\alpha\dot\alpha}, \mu^{-2} f,  \mu^{-2} \bar f),
\ee where $\mu$ is a real parameter. As we show below, the
transformation \eqref{bhussym} with $\mu=-1$, that does not affect
$f$, describes the $\tau$-symmetry transformation \eqref{tau}.

The equations \eqref{feq} and \eqref{feqc}
result from the Bianchi identities for
the curvatures \eqref{FDA3}, \eqref{FDA3c}.  It can be shown that
$f(X)=f(\gk^2)$ as a consequence of \eqref{FDA2}. Indeed, we have
\be \label{Dkap2}
d\kappa^2 = (\lambda^2+\frac13
f\kappa^2)\fh^{\alpha\dot\alpha}\fV^{\alpha}{}_{\dot\alpha}\kappa_{\alpha\alpha}.
\ee
Comparing \eqref{feq} and \eqref{Dkap2}, we obtain
\be \label{diff}
\frac{d\kappa^2}{2\lambda^2+\frac23
f\kappa^2}+\frac{df}{\frac16f^2+5\lambda^2\frac{f}{\kappa^2}} = 0.
\ee
Its general solution is
\be \label{f-solve}
f=6\M\frac{\G^3}{\kappa^2},
\ee
where the complex parameter $\M$ appears as an integration constant
and $\G$ is defined implicitly
by
\be \label{Gdef}
\M \G^3 -\G\sqrt{-\kappa^2}= \lambda^2
\ee
and satisfies
\be \label{dG}
d\G=-\frac{\G^2}{2\sqrt{-\kappa^2}}
\fh^{\alpha\dot\alpha}\fV^{\alpha}{}_{\dot\alpha}\kappa_{\alpha\alpha}.
\ee

Analogously,
\be \label{f-solvec}
\bar f=6\bar \M\frac{\bar \G^3}{\bar\kappa^2}\q
\bar \M \bar\G^3 -\bar\G\sqrt{-\bar\kappa^2}= \lambda^2.
\ee

\subsection{Killing projectors} \label{S:KP}

Let two pairs of mutually conjugated projectors
$\Pi^{\pm}_{\al\gb}$ and $\bar\Pi^{\pm}_{\dal\dgb}$ have the form
\be \label{introP}
\Pi^\pm_{\alpha\beta} = \frac12 (\epsilon_{\alpha\beta}  \pm \frac{1}{\sqrt{-\kappa^2}}\kappa_{\alpha\beta}), \qquad
\bar\Pi^\pm_{\dot\alpha\dot\beta} = \frac12
(\epsilon_{\dot\alpha\dot\beta}  \pm
\frac{1}{\sqrt{-\bar\kappa^2}}\bar\kappa_{\dot\alpha\dot\beta}).
\ee
They satisfy
\be
\Pi^\pm_{\alpha}{}^{\beta}\Pi^\pm_{\beta\gamma} =
\Pi^\pm_{\alpha\gamma}\,, \qquad
\Pi^\pm_{\alpha}{}^{\beta}\Pi^\mp_{\beta\gamma} = 0\q
\bar\Pi^\pm_{\dot\alpha}{}^{\dot\beta}\bar\Pi^\pm_{\dot\beta\dot\gamma}
= \bar\Pi^\pm_{\dot\alpha\dot\gamma}\,, \qquad
\bar\Pi^\pm_{\dot\alpha}{}^{\dot\beta}\bar\Pi^\mp_{\dot\beta\dot\gamma}
= 0\,.
\ee
From the definition \eqref{introP} it follows  that
\be
\Pi^\pm_{\alpha\beta} = - \Pi^\mp_{\beta\alpha}\,, \qquad
\bar\Pi^\pm_{\dot\alpha\dot\beta} = -
\bar\Pi^\mp_{\dot\beta\dot\alpha}\,.
\ee
Hereinafter we will focus on the holomorphic ({\it i.e.}, undotted)
sector of the BHUS. All relations in the antiholomorphic sector
result by conjugation.

From \eqref{introP} and \eqref{FDA1} and \eqref{FDA2} it follows
that
\be \label{derPi}
\D\Pi^\pm_{\alpha\al} = \pm \G
(\Pi^{+}_{\al\gb}\Pi^{+}_{\al\gb}+\Pi^{-}_{\al\gb}\Pi^{-}_{\al\gb})\fV^{\gb}{}_{\dgga}\fh^{\gb\dgga}.
\ee
The projectors \eqref{introP} split the two-dimensional
(anti)holomorphic spinor space into the direct sum of two
one-dimensional subspaces. For any $\xi_\alpha$ we set
\be
\xi^\pm_{\alpha} = \Pi^\pm_{\alpha}{}^{\beta} \xi_{\beta}\,,
\qquad \xi^+_\alpha+\xi^-_\alpha = \xi_\alpha,
\ee
so that
$
\Pi^\mp_{\alpha}{}^{\beta} \xi^\pm_{\beta} = 0\,.
$
This allows us to build light-like vectors with the aid of
projectors. Indeed, consider an arbitrary vector
$U_{\alpha\dot\alpha}$. Using \eqref{introP} define
$U^{\pm}_{\alpha\dot\alpha}$ and
$U^{\pm\mp}_{\alpha\dot\alpha}$ as
\be
U^{\pm}_{\alpha\dot\alpha} = \Pi^\pm_{\alpha}{}^{\beta}
\bar\Pi^\pm_{\dot\alpha}{}^{\dot\beta} U_{\beta\dot\beta}\q
U^{+-}_{\alpha\dot\alpha} =  \Pi^+_{\alpha}{}^{\beta}
\bar\Pi^-_{\dot\alpha}{}^{\dot\beta} U_{\beta\dot\beta}\,, \qquad
U^{-+}_{\alpha\dot\alpha} =  \Pi^-_{\alpha}{}^{\beta}
\bar\Pi^+_{\dot\alpha}{}^{\dot\beta} U_{\beta\dot\beta}\,.
\ee
Obviously,
$U^{\pm}_{\alpha\dot\beta}U^{\pm\alpha\dot\gamma} = 0$ and
$U^{\pm}_{\alpha\dot\alpha}U^{\pm\beta\dot\alpha} = 0$. Then
$U^{-}_{\alpha\dot\alpha}$ can be cast into the form
\be \label{uspin}
U^{-}_{\alpha\dot\alpha} = \psi_{\alpha} \bar\zeta_{\dot\alpha}\,,
\ee
From this it follows that
\be
U^{\pm}_{\alpha\dot\beta}U^{\pm}_{\beta\dot\alpha} =
U^{\pm}_{\alpha\dot\alpha}U^{\pm}_{\beta\dot\beta}\,, \qquad
U^{-+}_{\alpha\dot\beta}U^{+-}_{\beta\dot\alpha} =
-\frac{(U^{-+}U^{+-})}{(U^{-}U^+)}
U^{-}_{\alpha\dot\alpha}U^{+}_{\beta\dot\beta}\,,
\ee
where $(AB) = A_{\alpha\dot\alpha}B^{\alpha\dot\alpha}$.

\subsection{Kerr-Schild vector in  BH unfolded system} \label{S:ngc}

Let us identify $\gk_{\al\al}$ in \eqref{introP} with
$\gk_{\al\al}$ in the BHUS and introduce two null vectors
\be \label{k-def}
k_{\alpha\dot\alpha} = \frac{2}{(\fV^{-}\fV^+)}\fV^{-}_{\alpha\dot\alpha}, \qquad
n_{\alpha\dot\alpha} = \frac{2}{(\fV^{-}\fV^+)}\fV^{+}_{\alpha\dot\alpha}
\ee
with the evident property
$
\f12k_{\al\dal}\fV^{\al\dal}=\f12n_{\al\dal}\fV^{\al\dal}=1.
$
It is a matter of definition which of two vectors
$k_{\al\dal}$ or $n_{\al\dal}$ to identify with a Kerr-Schild vector.
Indeed, the equations (\ref{FDA1})--(\ref{feqc}) are
invariant under the symmetry $\tau_{-1}$ \eqref{bhussym} which acts
on $k_{\alpha\dot\alpha}$ as
$\tau_{-1}( k_{\al\dal})= -n_{\al\dal}\,.$
Let us choose $k_{\alpha\dot\alpha}$ as a Kerr-Schild vector.

Obviously,
\be \label{null-cond}
k_{\alpha\dot\alpha}k^{\beta\dot\alpha}=0\,.
\ee
From BHUS and projector properties the geodesity condition
follows by straightforward calculation
\be \label{congr}
k^{\alpha\dot\alpha}\fD_{\alpha\dot\alpha} k_{\beta\dot\beta}=0\,.
\ee
In addition, $k_{\alpha\dot\alpha}$ is the eigenvector of the Papapetrou field $\kappa_{\alpha\alpha}$
\be
\kappa_{\alpha\beta}k^{\beta}{}_{\dot\alpha}=\sqrt{-\kappa^2}
k_{\alpha\dot\alpha}\,
\ee
and has the following
properties as a consequence of \eqref{FDA1}, \eqref{FDA2} and \eqref{dG}
\be\label{prop1}
\fD_{\alpha\dot\alpha} k^{\alpha\dot\alpha}=-2(\G+\bar{\G})\,, \qquad
k^{\alpha\dot\alpha}\fD_{\alpha\dot\alpha}\G=2\G^2\,,
\ee
\be\label{prop2}
\fD_{\alpha\dot\alpha}
((\G+\bar{\G})k^{\alpha\dot\alpha})=-4\G\bar{\G}\,,\qquad
\fD_{\alpha\dot\alpha} (\G\bar{\G}k^{\alpha\dot\alpha})=0\,,
\ee
\be \label{prop3}
k_{\alpha}{}^{\dot\alpha} \fD_{\alpha\dot\alpha} k_{\gamma\dot\gamma}
= \G  k_{\alpha}{}^{\dot\alpha} \fV_{\gamma\dot\alpha} k_{\alpha\dot\gamma}.
\ee

The BHUS \eqref{FDA1}--\eqref{FDA4} admits a sourceless Maxwell
tensor. Indeed, consider
$F_{\alpha\alpha}$ of the form
\be \label{maxwell}
F_{\alpha\alpha}=\frac{\G^2}{\sqrt{-\kappa^2}}\kappa_{\alpha\alpha}\,.
\ee
That \eqref{maxwell} solves Maxwell equations can be easily
verified as follows. Using \eqref{Gdef}, \eqref{k-def} and
\eqref{FDA1}, \eqref{FDA2} one can make sure that
\be \label{max-dif}
F_{\alpha\alpha}=\frac12 \fD_{\alpha\dot\alpha}
((\G+\bar{\G})k_{\alpha}{}^{\dot\alpha})
\ee
and $F_{\alpha\alpha}$ is $\tau_{-1}$-invariant. In other words, 
the vector-potential $A_m=\frac12(\G+\bar{\G})k_m$
gives the Maxwell tensor field $F=dA$.

Due to \eqref{maxwell}, the Weyl tensor \eqref{weyl} can be
rewritten in the  form
\be \label{weyl-max}
C_{\al(4)}= -\frac{6\M}{\G}F_{\alpha\alpha}F_{\alpha\alpha}\,.
\ee
The differentiation of \eqref{maxwell} with the aid of
\eqref{FDA2} yields
\be \label{dif-F}
\M \fD F_{\alpha\alpha} = \frac14
\fh^{\beta\dot\beta}\fV^{\beta}{}_{\dot\beta}C_{\beta\beta\alpha\alpha}\,.
\ee
The Maxwell equations
\be \label{max+bian}
\fD_{\gamma\dot\alpha}F_{\alpha}{}^{\gamma} = 0\,, \qquad
\fD_{\alpha\dot\gamma}\bar{F}_{\dot\alpha}{}^{\dot\gamma} = 0\,
\ee
are now  simple consequences of \eqref{dif-F}.

The Maxwell tensor \eqref{maxwell} is related to Killing-Yano tensor  via
\be \label{yka}
Y_{\al\al}=\f{i}{\G^3}F_{\al\al}\,.
\ee
Indeed, \eqref{yka} satisfies
\be \label{yano}
\D_{\al\dal}Y_{\al\al}=0\,,\quad
\D_{\gb\dal}Y^{\gb}{}_{\al}+\D_{\al\dgb}\bar{Y}^{\dgb}{}_{\dal}=0\,,
\ee
which implies the Yano-Killing equation \cite{Yano}
$\D_{(k}Y_{m)n}=0$ for  $Y_{mn}=-Y_{nm}$.

As follows from \eqref{f-solve}, \eqref{Gdef},
\eqref{maxwell} and \eqref{yka}, $Y_{\al\al}$ and $F_{\al\al}$ constitute the
Papapetrou two-form
\be \label{PFY-fda}
\gk_{\al\al}=\M F_{\al\al}+i\gl^2Y_{\al\al} = \left(\M-\frac{\lambda^2}{G^3}\right)F_{\alpha\alpha}.
\ee

\section{Kerr black hole unfolded system from $AdS_4$}\label{Kerr}

As a preparation to the description of the Kerr-Schild Ansatz in
BHUS, let us reformulate the $AdS_4$ geometry in spinor notation.
The equations \eqref{exmp}--\eqref{zero-2} are equivalent
to the following special case of the BHUS \eqref{FDA1}--\eqref{FDA4} with $f=0$ that describes the vacuum
solution with vanishing Weyl tensor, {\it i.e.}, empty $AdS_4$ space
\begin{eqnarray}
DV_{\alpha\dot\alpha} & =& \frac12
h^{\gamma}{}_{\dot\alpha}\kappa_{0\gamma\alpha}+\f12h_{\al}{}^{\gga}\bar{\gk}_{0\dal\dgga}\,,
 \label{FDA1-ads} \\
D\kappa_{0\alpha\alpha} &=&\lambda^2
h_{\alpha}{}^{\dot\gamma}V_{\alpha\dot\gamma}\,, \label{FDA2-ads} \\
D\bar\kappa_{0\dot\alpha\dot\alpha} &=&\lambda^2
h^{\gamma}{}_{\dot\alpha} V_{\gamma\dot\alpha}\,, \label{FDA2c-ads} \\
Dh_{\alpha\dot\alpha}&=&0\,,\\
R_{\alpha\alpha} &=&
d\Omega_{\alpha\alpha} + \Omega_{\alpha}{}^{\beta} \wedge \Omega_{\beta\alpha} =
\f{\lambda^2}{2}h_{\alpha\dot\alpha}
\wedge h_{\alpha}{}^{\dot\alpha}\,,\label{FDA3-ads}\\
\bar R_{\dot\alpha\dot\alpha} &=&
d\bar\Omega_{\dot\alpha\dot\alpha} + \bar\Omega_{\dot\alpha}{}^{\dot\beta} \wedge \bar\Omega_{\dot\beta\dot\alpha} =
 \f{\lambda^2}{2}h_{\alpha\dot \alpha}
\wedge h^{\alpha}{}_{\dot\alpha}\,,\label{FDA4-ads}
\end{eqnarray}
where  $h_{\al\dal}$ is the $AdS_4$ vierbein,
$\Omega_{\alpha\alpha}$ and $\bar\Omega_{\dot\alpha\dot\alpha}$ are
components of Lorentz connection, $D$ is the
background Lorentz differential, $V_{\al\dal}$ is an $AdS_4$ Killing
vector and $R_{\alpha\alpha}$, $\bar R_{\dot\alpha\dot\alpha}$ are the components of $AdS_4$ curvature
two-form:
$
D^2\xi_{\alpha\dot\alpha}=\frac12 R_{\alpha}{}^{\beta}\xi_{\beta\dot\alpha}
+ \frac12 \bar R_{\dot\alpha}{}^{\dot\beta}\xi_{\alpha\dot\beta}.
$

The $AdS_4$ Killing projectors $\Pi^\pm_{0\alpha\beta}$
single out the null vector $k_{0\alpha\dot\alpha}$ that defines $AdS_4$
null geodesic congruence
\be \label{congr-ads}
k_0^{\alpha\dot\alpha}D_{\alpha\dot\alpha}
k_{0\beta\dot\beta}=0\,,
\ee
where
\be \label{k-vac}
k_{0\alpha\dot\alpha} =
\frac{2}{(V^{-}V^+)}V^{-}_{\alpha\dot\alpha}\,.
\ee
Analogously, the Maxwell field generated by the null vector
$k_{0\alpha\dot\alpha}$ is defined by
\be \label{F-vac}
F_{0\alpha\alpha}=-\lambda^{-2}\G_0^3\kappa_{0\alpha\alpha}\,,
\ee
with
\be \label{G-ads}
\G_0=-\frac{\lambda^2}{\sqrt{-\kappa_0^2}}\,.
\ee

Now we are in a position to show that the $AdS_4$ Kerr BH is a solution
of BHUS \eqref{FDA1}--\eqref{FDA4} resulting
at real $\M$ from some algebraic field redefinition
\be\label{trans}
(h_{\al\dal}, V_{\al\dal}, \gk_{0\al\al})\to (\fh_{\al\dal},
\fV_{\al\dal}, \gk_{\al\al})\,,
\ee
that, in fact, extends
 the Kerr-Schild Ansatz to the full BHUS, expressing the
$AdS_4$ Kerr BH  entirely in terms of the
$AdS_4$ background unfolded system.

Let the sets of
fields $(\fO_{\al\al}, \fh_{\al\dal}, \fV_{\al\dal},
\gk_{\al\al})$ and $(\Omega_{\al\al}, h_{\al\dal}, V_{\al\dal},
\gk_{0\al\al})$ be the Lorentz connections, vierbeins, Killing
vectors and Papapetrou fields of the BHUS
\eqref{FDA1}--\eqref{FDA4}
and of the vacuum equations
\eqref{FDA1-ads}--\eqref{FDA4-ads}, respectively. The explicit
map between these two unfolded system is
\be\label{final omega}
\fO_{\al\dal|\gga\gga}=\Omega_{\al\dal|\gga\gga}-\frac{2\M\G_{0}^2}{\sqrt{-\kappa_{0}^2}}\kappa_{0\gamma\gamma}k_{0\alpha\dot\alpha}
-\frac{\M\G_0}{2}(\G_0+\bar \G_0)
k_{0\gamma}{}^{\dot\gamma}V_{\alpha\dot\gamma}k_{0\gamma\dot\alpha}\,,
\ee
\be \label{KS-shift}
\fh_{\alpha\dot\alpha}=h_{\alpha\dot\alpha} +
\frac{\M}{4}(\G_0+\bar \G_0)
h^{\beta\dot\beta}k_{0\beta\dot\beta}k_{0\alpha\dot\alpha}\,,
\ee
\begin{eqnarray} \label{V-rel}
\fV_{\alpha\dot\alpha} &=&V_{\alpha\dot\alpha} + \frac{\M}{2}(\G_0 + \bar \G_0)k_{0\alpha\dot\alpha}\,, \\
\kappa_{\alpha\alpha}&=& (1-\M\lambda^{-2}\G_{0}^3)
\kappa_{0\alpha\alpha}\,. \label{kappa-rel}
\end{eqnarray}
As a consequence of \eqref{KS-shift}, the metric has the
Kerr-Schild form
\be\label{Kerr-Schild}
g_{mn}=\eta_{mn}+\M(\G_0+\bar{\G}_0)k_{0m}k_{0n}.
\ee
Another consequence is the invariance  of the null congruence
\eqref{k-vac} and function $\G_0$ under the deformation
\be\label{inv}
\G_0=\G\,,\qquad k_{0\al\dal}=k_{\al\dal}\,.
\ee
Note that  \eqref{inv} along with \eqref{max-dif} entail the
equality $F_{0\al\al}=F_{\al\al}$  and thus determine the Weyl
tensor \eqref{weyl-max} in terms of $AdS_4$ geometry.

Let us sketch the main steps of the derivation of the relations
\eqref{final omega}--\eqref{kappa-rel} in some more detail.
To see that
\eqref{final omega}--\eqref{kappa-rel} indeed relate two unfolded
systems and to simplify the subsequent analysis, we start with the
vacuum system \eqref{FDA1-ads}--\eqref{FDA4-ads} and suppose that
the deformation is geodesic, {\it i.e.}, \eqref{trans}
leaves the Kerr-Schild vector and
the function $\G_0$ invariant \eqref{inv}, checking this afterwards.

Let us look for the Kerr-Schild deformation of the background vierbein field
\be \label{KS-shift1}
\fh_{\alpha\dot\alpha}=h_{\alpha\dot\alpha} + \frac{\M}{4}(\G+\bar
\G) h^{\beta\dot\beta}k_{\beta\dot\beta}k_{\alpha\dot\alpha}\,.
\ee
Imposing the zero-torsion condition on the deformed vierbein
$\fD\fh_{\alpha\dot\alpha}=0$,
one obtains  the deformed Lorentz connection $\fO_{\alpha\alpha} =
\Omega_{\alpha\alpha}+\omega_{\alpha\alpha}$ with
\be
\omega_{\alpha\dot\alpha|\gamma\gamma} =
-\fh^i{}_\gamma{}^{\dot\gamma}D_{\gamma\dot\gamma}
\fh_{i\alpha\dot\alpha}\,.
\ee
It follows then that
\be \label{diff-omega}
\omega_{\alpha\dot\alpha|\gamma\gamma} =
-\frac{\M}{2}D_{\gamma\dot\gamma}
((\G+\bar{\G})k_{\gamma}{}^{\dot\gamma}k_{\alpha\dot\alpha}).
\ee
Using \eqref{null-cond} and \eqref{congr} one observes that the
background Lorentz derivative $D$ in \eqref{diff-omega}
can be replaced by the deformed one $\fD$. Using \eqref{FDA1}, 
\eqref{FDA2} and \eqref{dG} we get \eqref{final omega}.

Now it is straightforward to check that the Weyl tensor admits the
following representation
\be
C_{\alpha\alpha\alpha\alpha} =
\fD_{\alpha\dot\alpha}\omega_{\alpha}{}^{\dot\alpha}{}_{|\alpha\alpha}.
\ee
Note that here $\D$ can again be replaced by $D$, {\it i.e.}, the terms
quadratic in $\omega$ cancel.

The substitution of $h_{\al\dal}$ \eqref{KS-shift1} and
$\Omega_{\alpha\alpha} = \fO_{\alpha\alpha}-\omega_{\alpha\alpha}$
into the $AdS_4$ unfolded system \eqref{FDA1-ads}--\eqref{FDA4-ads}
yields BHUS \eqref{FDA1}--\eqref{FDA4}
with the Killing vector and the Killing two-form transformed according to
\eqref{V-rel} and \eqref{kappa-rel}. Finally, it is not hard to verify
that \eqref{inv} is consistent with \eqref{V-rel}, \eqref{kappa-rel}.

Thus, the BHUS that describes the $AdS_4$ BH geometry with non-trivial Weyl
tensor results from the algebraic field redefinition of the
$AdS_4$ vacuum vierbein and connection, providing Kerr-Schild type
vacuum solution of Einstein equations.

The identification with the standard BH description of Section \ref{genfacts}
requires
\be \label{QGM}
\G= \frac1Q\,, \quad \mathcal{M} = M,
\ee
with $Q$ \eqref{Qdef} and  the Killing vector $\fV^i$
\eqref{Kill}.
Then the Kerr-Schild vector $k^i$ determined from
\eqref{k-vac} coincides with \eqref{k-comp}, whereas
$n^i$, which arises with the aid of the discrete symmetry $\tau_{-1}$,
coincides with one from Section \ref{genfacts}.
Moreover, the metric \eqref{Kerr-Schild} is identical to standard
Kerr-Schild representation \eqref{ks}.

The Schwarzschild case with $a=0$ is singled out by the
following additional condition
\be
\fV^{+-}_{\al\dal}=\fV^{-+}_{\al\dal}=0\,.
\ee

\section{Black hole  massless fields}\label{HS}

Let us show how the  BHUS reproduces
solutions  to free massless field equations in
$AdS_4$ for all integer spins via a  Kerr-Schild type
algebraic field redefinition.

Consider the traceless symmetric tensor
\be
\varphi_{mm}=\frac12(\G+\bar \G)k_m k_m\,.
\ee
Taking into account \eqref{FDA1}, \eqref{FDA2} and \eqref{FDA3},
the straightforward calculation yields
\be\label{ein2}
\fD^n\fD_n \varphi_{mm} - 2\fD^n \fD_m
\varphi_{mn} =-6\lambda^2 \varphi_{mm}.
\ee
In this way we obtain Einstein equations for the Kerr-Schild
decomposition \eqref{Kerr-Schild}, or spin-2 free field equations.
Note that one can use $D$ instead of $\D$ in \eqref{ein2}.

The Maxwell case is analogous. It is a simple consequence
of \eqref{max-dif} that the vector field $\varphi_m=\frac12(\G+\bar \G)k_m$ satisfies
\eqref{max+bian}.

Things change for the scalar $\varphi=\frac12(\G+\bar{\G})$. Using
\eqref{FDA1}, \eqref{FDA2}, \eqref{dG} and \eqref{max+bian} one finds
\be
\fD^{\alpha\dot\alpha}\fD_{\alpha\dot\alpha}\G=-4\G^2\sqrt{-\kappa^2}\,.
\ee
The substitution of \eqref{Gdef} yields
\be\label{bh-scalar}
\fD^{\alpha\dot\alpha}\fD_{\alpha\dot\alpha}\G=4\lambda^2 \G -4\M\G^4.
\ee
Thus, it reduces to
$
D^{\alpha\dot\alpha}D_{\alpha\dot\alpha}\varphi = 4
\lambda^2\varphi
$
only in the $AdS_4$ limit with $\M=0$, $\fD=D$. This equation describes
propagation of a massless scalar in $AdS_4$.
Using differential properties of $k_{\alpha\dot\alpha}$, one can obtain
\be
\fD_m(k^mk^n \fD_n(\G+\bar \G)^2) = 4(\G^4+\bar{\G}^4)\,.
\ee
Hence, from the equation \eqref{bh-scalar} it follows
\be
\fD^m \fD_m\varphi=2\lambda^2\varphi -2 \M
\fD_m(\varphi^{mn} \fD_n\varphi)\,.
\ee
The second term on the r.h.s. represents the nonlinear
correction to a massless scalar propagating in Kerr-Schild
background. Thus, the fields $\varphi, \varphi_m, \varphi_{mm}$
verify the free massless
equations in $AdS_4$ space   for spins 0,1 and 2,
respectively, \cite{CF,Mets}. Using BHUS one can show that\footnote{A number
in parenthesis next to an index denotes a number of symmetrized
indices, e.g., $\phi_{m(s)} =\phi_{m_1\ldots m_s}$.}
\be
\varphi_{m(s)}=\frac12(\G+\bar{\G})k_{m}\dots k_{m}
\ee
gives a Kerr-Schild solution for a massless integer spin--$s$
equation in $AdS_4$ background
\be\label{free-s}
D^n D_n \varphi_{m(s)}-sD_n D_m \varphi^{n}{}_{m(s-1)}=-2(s-1)(s+1)\gl^2\varphi_{m(s)}\,.
\ee
In the  Kerr-Schild background \eqref{free-s}  changes to
\be\label{s-int}
\D^n \D_n \varphi_{m(s)}-s\D_{n}\D_{m}\varphi^{n}{}_{m(s-1)}=-2(s-1)(s+1)\gl^2\varphi_{m(s)}
-\M(s-1)(s-2)\D_n (\varphi^{nr}\D_{r}\varphi_{m(s)})\,.
\ee
Note that the interaction term in \eqref{s-int} vanishes only for
$s=1$ and $s=2$.

This result suggests that the Kerr-Schild Ansatz  should admit
an extension to the nonlinear equations of $4d$
massless fields of
all spins  (see \cite{Gol} and references therein).

\section{Summary and discussion}
\label{sum}

We have shown that $AdS_4$ Kerr BH admits a simple
description in terms of unfolded field equations generated by
a Killing vector of the background $AdS_4$ space.
We considered the case of four dimensions using the spinor formalism.

Our aim was to find the description of the $AdS_4$ Kerr BH that
does not refer to a particular coordinate system. Such a description
is given in terms of a coordinate-independent unfolded system
of differential equations that encode all properties of
the $AdS_4$ Kerr BH.

The proposed approach allowed us to shown how seemingly
different structures resided in the $4d$ BH such as, e.g.,
the existence of
Yano-Killing tensor, the decomposition of the curvature tensor
into Maxwell tensor and the Papapetrou field,  naturally
arise from the $AdS_4$ geometry with a distinguished
Killing vector. An interesting direction for the further study is to explore
the BH unfolded system on its own right, including, in particular, the case
of complex deformation parameter $\M$ which, as we expect, should
correspond to the Taub-NUT case. More generally, such an
analysis can provide a powerful technique for
identification of BH solutions via the algebraic properties of the
corresponding solutions.

We believe that the proposed construction  allows an extension
to the nonlinear HS field equations also formulated in the
unfolded form (see \cite{Gol,SSS,solv} for reviews). It is worth
to note that the proposed BH unfolded system
allows us to build solutions not only to linearized
Einstein equations which reduce in
this case to Pauli-Fierz equations but also to Fronsdal equations of
all massless integer spins propagating in $AdS_4$. This fact
looks particularly  encouraging  from the HS gauge theory  perspective.
Indeed, this implies that $AdS_4$ Kerr BH
naturally fits linearized HS gauge theory through an
algebraic field redefinition of the vacuum solution.
The analysis of the BH solution in the nonlinear HS gauge theory
is, however, beyond the scope of this paper and will be given elsewhere.

\section*{Acknowledgement}

This research was supported in part by INTAS Grants No 03-51-6346
and  05-7928, RFBR Grant No 05-02-17654, LSS No 4401.2006.2. A.M.
acknowledges financial support from Landau Scholarship Foundation
and from Dynasty Foundation.

\end{document}